\documentclass[11pt]{article}
\usepackage{epsf}
\usepackage{color}
\usepackage{amsmath,amssymb,color,graphics,amscd,amsfonts}

\begin{document}

\begin{flushright} 
August, 2005  \\
 KUNS-1964 \\
 %RIKEN-TH-??\\
 % \today \\
\end{flushright} 

\vspace{0.1cm}

\begin{Large}
\vspace{1cm}
\begin{center}
{\bf Describing Curved Spaces by Matrices} \\ 
\end{center}
\end{Large}

\vspace{1cm}

\begin{center}
{\large Masanori Hanada$^{a}$, Hikaru Kawai$^{ab}$ 
and Yusuke Kimura$^{a}$ }   \\ 

\vspace{0.5cm} 
$a)$ Department of Physics, Kyoto University, 
Kyoto 606-8502, Japan \\
$b)$ Theoretical Physics Laboratory, RIKEN, Wako 351-0198, 
Japan \\

\vspace{0.5cm} 
{\sf hana, hkawai, ykimura@gauge.scphys.kyoto-u.ac.jp}
\vspace{0.8cm} 

\end{center}

\vspace{1cm}

\begin{abstract}
\noindent
\end{abstract}

It is shown that a covariant derivative on any $d$-dimensional 
manifold $M$ can be mapped to a set of $d$ operators acting on the 
space of functions defined 
on the principal $Spin(d)$-bundle over $M$.
In other words, any $d$-dimensional manifold can be described in terms 
of $d$ operators acting on an infinite-dimensional space.
Therefore it is natural to introduce a new interpretation of matrix 
models in which matrices represent such operators.
In this interpretation, the diffeomorphism, local Lorentz symmetry 
and their higher-spin analogues are included in the unitary symmetry 
of the matrix model. 
Furthermore, the Einstein equation is obtained from the equation of 
motion, if we take the standard form of the action, 
$S=-tr\left([A_{a},A_{b}][A^{a},A^{b}]\right) $.

\newpage

\section{Introduction}
\hspace{0.51cm}
Although it is believed that 
string theory may provide 
the unification of fundamental interactions, 
its present formulation based on perturbation theory is not satisfactory. 
In order to examine whether it really describes our four-dimensional world,
a non-perturbative and background independent formulation is needed. 
Some of the promising candidates are matrix models \cite{BFSS,IKKT}. 
They are basically obtained through dimensional reduction of 
the ten-dimensional $U(N)$ ${\cal N}=1$ supersymmetric Yang-Mills 
theory. 

The action of IIB matrix model \cite{IKKT} is given by 
\begin{eqnarray}
  S=-\frac{1}{4g^{2}}tr\left([A_{a},A_{b}]
    [A^{a},A^{b}]\right) 
    +\frac{1}{2g^{2}}tr\left(
  \bar{\psi}\Gamma^{a}[A_{a},\psi]\right), 
  \label{IIBaction}
\end{eqnarray}
where $\psi$ is a ten-dimensional Majorana-Weyl spinor, and 
$A_{a}$ and $\psi$ are $N\times N$ hermitian matrices. 
The indices $a$ and $b$ are 
contracted by the flat metric. 
This action has an $SO(10)$ global Lorentz symmetry and $U(N)$ symmetry. 
There is evidence that this model describes gravity
through a one-loop quantum effect 
\cite{IKKT,hep-th/0503101}. 
However, it is unclear how 
the fundamental principle of general relativity 
is realized in this model. 
In order to elucidate this point, we need to clarify 
the meaning of the flat metric. This should help us to 
understand the dynamics of this model further. 

\vspace{0.2cm}

Let us first point out that this model has several interpretations. 
First, this action can be regarded as the Green-Schwarz action of 
IIB superstring in the Schild gauge 
after the regularization of the  
functions on the world-sheet by the matrices \cite{IKKT}. 
In this case, the matrices $A_{a}$ are simply the
coordinates of the target space.\footnote{
This action has the same form as 
the low energy effective action of D-instantons 
in the flat background \cite{hep-th/9510135}, and 
the eigenvalues of the matrices represent the positions of 
the D-instantons. Here, we do not take this point of view.
}
This interpretation 
is consistent with the supersymmetry algebra. 
If we regard a constant shift of matrices as a translation, 
we have the ${\cal N}=2$ supersymmetry. 

Another interpretation of this model is based on
noncommutative geometry.
The matrix model 
(\ref{IIBaction}) has the following 
noncommutative momenta as a classical solution, 
\begin{equation}
A_{a}=p_{a}, \quad [p_{a},p_{b}]=iB_{ab}1, 
\end{equation}
where $B_{ab}$ is a $c$-number.  
Then we introduce noncommutative coordinates $x^{a}$ as 
\begin{equation}
x^{a}=C^{ab}p_{b}, \quad C^{ab}=(B^{-1})^{ab}.
\label{NCcoordinatefromp}
\end{equation}
These coordinates satisfy the following commutation relations:
\begin{equation}
[x^{a},x^{b}]=-iC^{ab}1, \quad
[x^{a},p_{b}]=i\delta^{a}{}_{b}1.
\label{noncommutativerelationforp}
\end{equation}
As is seen from this, there is no essential difference between coordinates and momenta
in noncommutative geometry. 
If we expand $A_{a}$ around this background as
\begin{equation}
A_{a}=p_{a}+a_{a}(x), 
\label{expansionofmatrixaroundNCbg}
\end{equation}
$a_{a}(x)$ becomes a gauge field on the noncommutative space 
\cite{hep-th/9612222,hep-th/9711162,AIIKKT}, and
$A_{a}$ can be regarded as the covariant derivative itself.

In the first interpretation presented above, the 
matrices represent the spacetime coordinates, while in the second 
they are regarded as differential operators
in a noncommutative space. 
Therefore it is natural to consider a third interpretation 
in which matrices represent differential operators 
in a commutative space. 
As a first attempt to realize such an interpretation, 
we regard large $N$ matrices as a linear mapping 
from $W$ to $W$, where $W$ is the space of smooth functions from 
${\mathbb R}^{10}$ to ${\mathbb C}$. In other words, $W$ is the set of 
field configurations of a scalar field 
in ten-dimensional space-time. 
Therefore matrices are identified with integral kernels, 
which can formally be expressed as 
differential operators of infinite order as 
\begin{eqnarray}
f(x) &\rightarrow &\int d^{10}y K(x,y)f(y) \cr
&\simeq &\left(b(x)+b^{\mu}(x)\partial_{\mu}
+b^{\mu\nu}(x)\partial_{\mu}\partial_{\nu}+\cdots \right)f(x). 
\end{eqnarray}
In this way, matrices can be naturally regarded 
as differential operators, and an 
infinite number of local fields appear as 
the coefficients of derivatives. 

In this interpretation, 
\begin{equation}
A_{a}=i\delta_{a}{}^{\mu}\partial_{\mu}
\end{equation}
is a classical solution,
\footnote{
Below, we regard the indices $a$ and $\mu$ 
as the local Lorentz and spacetime indices, respectively.  
This solution corresponds to the flat background, with 
$e_{a}^{\mu}=\delta_{a}^{\mu}$.
} 
because the equation of motion is 
given by $[A_{a},[A_{a},A_{b}]]=0$ when $\psi=0$. 
We then expand $A_{a}$ around this solution \cite{hep-th/0204078},  
obtaining 
\begin{eqnarray}
A_{a}=
a_{a}(x)+\frac{i}{2}\{a_{a}^{\mu}(x),\partial_{\mu}\}
+\frac{i^{2}}{2}\{a_{a}^{\mu\nu}(x),\partial_{\mu}\partial_{\nu}\}
+\cdots , 
\label{derivativeexpansionboson} 
\end{eqnarray}
where the coefficients 
$a_{a}^{\mu\nu\cdots}(x)$ 
are real totally symmetric tensor fields, and 
anti-commutators have 
been introduced to make each term hermitian. 
The unitary symmetry of this model is expressed as
\begin{equation}
\delta A_{a}=i[\Lambda,A_{a}],
\end{equation}
where 
the infinitesimal parameter $\Lambda$ can also be expanded as 
\begin{eqnarray}
\Lambda=\lambda(x)+\frac{i}{2}\{\lambda^{\mu}(x),\partial_{\mu}\}
+\frac{i^{2}}{2}\{\lambda^{\mu\nu}(x),\partial_{\mu}\partial_{\nu}\}
+\cdots . 
\label{unitarytrparameterexpansion}
\end{eqnarray}

Let us examine how each local field transforms 
in the case that 
$\Lambda=\frac{i}{2}\{\lambda^{\mu}(x),\partial_{\mu}\}$, which is expected 
to give the diffeomorphism. 
For the first two fields, we obtain 
\begin{eqnarray}
&&\delta a_{a}(x)
%=i[i\lambda^{\mu}(x)\partial_{\mu},a_{a}(x)]
=-\lambda^{\mu}(x)\partial_{\mu}a_{a}(x), \cr
&&\delta a_{a}^{\mu}(x)
%=i[i\lambda^{\mu}\partial_{\mu},a_{a}^{\nu}\partial_{\nu}]
=-\lambda^{\nu}(x)\partial_{\nu}a_{a}^{\mu}(x)
+a_{a}^{\nu}(x)\partial_{\nu}\lambda^{\mu}(x). 
\end{eqnarray}
From this, we find that 
$a_{a}(x)$ and $a_{a}^{\mu}(x)$ transform as a scalar and 
a vector, respectively. 
The $SO(10)$ index $a$ has no relation to the diffeomorphism. 
Therefore we attempt to 
interpret it as a local Lorentz index 
and regard $a_{a}^{\mu}(x)$ as something like the vielbein field. 
Then it is natural to replace 
ordinary derivatives with covariant derivatives as
\begin{eqnarray}
A_{a}\rightarrow
%\overset{?}{=}
a_{a}(x)+\frac{i}{2}\{a_{a}^{b}(x),\nabla_{b}\}
+\frac{i^{2}}{2}\{a_{a}^{bc}(x),\nabla_{b}\nabla_{c}\}
+\cdots. 
\label{naiveexpansion}
\end{eqnarray}
Here, $\nabla_{a}$ is a covariant derivative, given by
\begin{eqnarray}
\nabla_{a}=e_{a}^{\mu}(x)
\left(\partial_{\mu}+\omega_{\mu}^{bc}(x){\cal O}_{bc} \right),
\label{firstcovariantderivative}
\end{eqnarray}
where $e_{a}^{\mu}(x)$ and $\omega_{\mu}^{ab}(x)$ 
are the vielbein and spin connection, respectively. 
The operator ${\cal O}_{ab}$ is the Lorentz generator, 
which acts on 
Lorentz indices. 
Its explicit form depends on the bundles 
on which the covariant derivative operates.

There arise two difficulties 
in carrying out the replacement (\ref{naiveexpansion}). 
One is the gluing of coordinate patches 
on curved spaces.
In order to define a vector operator like (\ref{naiveexpansion}), 
we first define it on each patch, and then glue them using 
transition functions.
The latter procedure usually mixes the vector components.
Therefore, a vector operator cannot be realized by
simply giving a set of ten matrices.
Such a set would be identified with a set of ten scalar operators, 
not a vector operator.
The second problem arises when we consider 
the product of covariant derivatives. 
Suppose $A_{a}=i\nabla_{a}$. 
Let us consider the  $1,2$ component of $A_{a}A_{b}$, as an example. 
It is given by
\begin{eqnarray}
A_{1}A_{2}=-\nabla_{1}\nabla_{2}
&=&-\partial_{1}\nabla_{2}-\omega_{1}^{2c}(x)\nabla_{c} \cr
&=&i\partial_{1}A_{2}+i\omega_{1}^{2c}(x)A_{c}
\end{eqnarray}
The difficulty comes from the second term. 
Because the sum is taken over the index $c$, 
$A_{1}A_{2}$ is not simply 
a product of $A_{1}$ and $A_{2}$.  
This prevents us from directly 
identifying $\nabla_{a}$ with a matrix and considering naive products. 
In other words, $\nabla_{a}$ cannot be identified with a matrix component
by component. 
The main goal of this paper is to solve these two problems. 
We show that it is possible to express covariant derivatives 
in terms of matrices. 
In subsection \ref{sec:S2example}, we present 
concrete examples of covariant derivatives on 
a two-sphere and a two-torus in terms of two matrices. 

The organization of this paper is as follows. 
In the next section,  
we explain how covariant derivatives 
on any curved space can be described by matrices. 
The regular representation of the Lorentz group plays an 
important role in this description. 
We regard matrices 
as mappings from the function space on the principal $Spin(10)$ bundle 
to itself. 
Then we present the above-mentioned examples.
In section \ref{sec:matrixmodelandloallorentz}, 
we apply this idea to matrix models. 
It is shown that the Einstein equation 
can be obtained from the equation of motion of the matrix model. 
If we introduce a mass term, we obtain a cosmological constant. 
We then consider 
the expansion of matrices with respect to covariant derivatives, 
and examine 
the local fields that appear as its coefficients. 
We find that both the diffeomorphism and the local Lorentz symmetry 
are naturally included in the unitary symmetry of the matrix model. 
Section \ref{sec:conclusiondiscussion} is devoted 
to discussions. 
In Appendix \ref{sec:proofofregulardecomposition}, 
we give the proof of (\ref{irreducibledecompregular}). 
In Appendix \ref{sec:HigherSpin}, 
we discuss higher-spin fields.

%%%%%%%%%%%%%%%%%%%%%%%%%%%%%%%%%%%%%%%%%%%%%%%%%%%%%%%%%%%%%%%%%%%%%%%%%%%%%%%

\section{Differential Operators on Curved Spaces}
\label{sec:deformedmatrixmodel}
\hspace{0.51cm}
In order to express covariant derivatives in terms of matrices, 
the vector space on which they act should be sufficiently large.  
A covariant derivative maps a tensor field of rank $n$ to one 
of rank $(n+1)$. 
Thus, if a space is invariant under the actions of covariant derivatives,
it should contain at least tensor fields of any rank. 
%What we have to do is to find a good space. 
In this section, we show that we can indeed find such a good space, 
in which covariant derivatives are expressed as 
endomorphisms on it.
Here we use the term {\it endomorphism on $V$} in reference to 
a linear map from a vector space $V$ to itself, 
and the expression $End(V)$ to represent the set of such maps. 
We further show that the use of this space resolves 
both of the difficulties mentioned in the previous section. 

In subsection \ref{sec:preliminaries}, 
we consider the regular representation of a group 
and examine some important properties of a vector bundle
whose fiber is its representation space. 
We then apply it to the Lorentz group and give
a prescription for embedding 
covariant derivatives into matrices in subsection \ref{sec:S2example}. 

\subsection{Preliminaries: Regular representation}
\label{sec:preliminaries}
\hspace{0.51cm}
Let us begin by considering the space of smooth functions 
from a group $G$ to ${\mathbb C}$: 
\begin{equation}
V_{reg}=\{f: G \rightarrow {\mathbb C}\}.
\label{spaceofregularrep}
\end{equation}
We assume $G$ to be compact and later 
choose $G$ as $Spin(10)$ or $Spin^{c}(10)$.  
The action of an element $h$ of $G$ is given by 
\begin{equation}
(\hat{h}f)(g)=f(h^{-1}g).
\label{gactiononV}
\end{equation}
The space $V_{reg}$ is called 
{\it the regular representation}. 
This representation is reducible and is decomposed into irreducible 
representations as 
\begin{equation}
V_{reg}=\underset{r} \oplus(
\underbrace{ V_{r}\oplus \cdots \oplus V_{r}
}_{d_{r}}
), 
\label{irreducibledecompregular}
\end{equation}
where $V_{r}$ is the space of the irreducible representation $r$ 
of $G$, and $d_{r}$ is its dimension. 
This decomposition is proven 
in Appendix \ref{sec:proofofregulardecomposition}. 

The following interesting isomorphism holds for 
any representation $r$:
\begin{equation}
V_{r} \otimes V_{reg} \cong 
\underbrace{ V_{reg} \oplus \cdots \oplus V_{reg}
}_{d_{r}}. 
\label{isomorphicrelationofVDotimesV}
\end{equation}
This is the key equation for expressing covariant derivatives 
as endomorphisms. 
The concrete form of 
(\ref{isomorphicrelationofVDotimesV}) is as follows. 
Let $\Phi^{i}(g)$ be an element of $V_{r} \otimes V_{reg}$, 
where the index $i$ transforms as the representation $r$. 
That is, the action of $G$ is given by 
\begin{equation}
(\hat{h}\Phi^{i})(g)=R^{i}{}_{j}(h)\Phi^{j}(h^{-1}g),
\end{equation}
where $R^{i}{}_{j}(h)$ is the representation matrix
corresponding to $r$. 
Then the isomorphism 
(\ref{isomorphicrelationofVDotimesV}) 
is given by\footnote{
Note that $R^{i}{}_{j}$ and $R^{(i)}{}_{j}$ are the same quantity. 
However, we distinguish them, because the indices $i$ and $(i)$ 
obey different transformation laws. Specifically, 
$i$ is transformed by the action of $G$, while $(i)$ is not. 
}
\begin{equation}
\Phi^{(i)}(g)=R^{(i)}{}_{j}(g^{-1})\Phi^{j}(g),  
\label{isomorphism}
\end{equation}
where $i=1,\cdots,d_{r}$.
We can verify that each component of $\Phi^{(i)}(g)$ belongs 
to $V_{reg}$ as 
\begin{eqnarray}
(\hat{h}\Phi^{(i)})(g)
&=&R^{(i)}{}_{j}(g^{-1})(\hat{h}\Phi^{j})(g) \cr
&=&R^{(i)}{}_{j}(g^{-1})R^{j}{}_{k}(h)\Phi^{k}(h^{-1}g) \cr
&=&\Phi^{(i)}(h^{-1}g).  
\label{indicateGandRcommute}
\end{eqnarray} 
The index $(i)$ is not
transformed by the action of $\hat{h}$ 
but, rather, it merely labels the copies of $V_{reg}$ 
on the right-hand side of (\ref{isomorphicrelationofVDotimesV}). 

Here we emphasize that 
this isomorphism is constructed in such a way that 
the action of $G$ commutes with 
the conversion of indices between $(i)$ and $i$. 
If $\Phi(g)$ has some number of indices, 
each index can be converted in a similar manner.
For example, if  $\Phi(g)$ has two indices, we have 
\begin{equation}
\Phi^{(i)j}(g)=R^{(i)}{}_{k}(g^{-1})\Phi^{kj}(g), 
\end{equation}
where 
$\Phi^{(i)j}(g)$ is an element of 
$V_{r^{\prime}} \otimes V_{reg}$ for each $i=1,\cdots,d_{r}$, 
and 
$\Phi^{kj}(g)$ is an element of 
$V_{r} \otimes V_{r^{\prime}} \otimes V_{reg}$. 

\vspace{0.4cm}

So far we have considered the regular representation $V_{reg}$ of $G$. 
We next consider a fiber bundle over a manifold $M=\cup_{i} U_{i}$, 
where the fiber is $V_{reg}$ and the structure group is $G$. 
We denote this fiber bundle by $E_{reg}$. 
A global section of $E_{reg}$ 
is defined as 
a set of smooth maps from each coordinate patch $U_{i}$ to $V_{reg}$. 
Because an element of $V_{reg}$ is a function from $G$ to ${\mathbb C}$,
such map is simply a function from $U_{i}\times G$ to ${\mathbb C}$.
In order to be glued globally, 
they must satisfy the following condition
on each overlapping region $x\in U_{i}\cap U_{j}$:
\footnote{
We use the symbol $[i]$ to indicate quantities 
associated with $U_{i}$.}
\begin{equation}
\tilde{f}^{[j]}(x,g)=\tilde{f}^{[i]}(x,t_{ij}(x)g), 
\end{equation}
where $t_{ij}(x)$ is a transition function. 
We denote the set of sections by $\Gamma(E_{reg})$
\footnote{
From 
(\ref{irreducibledecompregular}), 
the set of sections is equivalent to 
the space of the field configurations 
\begin{equation}
\underset{r}\oplus  (
\underbrace{ {\cal V}_{r}\oplus \cdots \oplus {\cal V}_{r}
}_{d_{r}}
), 
\label{section2}
\end{equation}
where ${\cal V}_{r}$ is the space of 
fields on $M$, which 
transforms as the irreducible representation $r$ of $G$. 
}. 

We next consider the principal bundle $E_{prin}$ which is 
associated with $E_{reg}$,
and introduce the space of smooth functions on it:
\begin{equation}
C^{\infty}(E_{prin})=\{f:E_{prin} \rightarrow {\mathbb C}\}. 
\end{equation}
Here we show that $\Gamma(E_{reg})$ is isomorphic to 
$C^{\infty}(E_{prin})$. 
To define an element $f$ of $C^{\infty}(E_{prin})$,
we first introduce a function $f^{[i]}$ 
from $U_{i}\times G$ to ${\mathbb C}$ for each patch $U_{i}$. 
They should be related on each overlapping region of 
$U_{i}$ and $U_{j}$ as 
\begin{equation}
f^{[j]}(x,g)=f^{[i]}(x,t_{ij}(x)g). 
\end{equation}
This follows from the fact that 
$E_{prin}$ is constructed from the set of
$U_{i}\times G$ 
by identifying
$(x_{[i]},g_{[i]})$ and $(x_{[j]},g_{[j]})$
on each overlapping region,
when they satisfy the relations 
\begin{equation}
x_{[i]}=x_{[j]}, \quad g_{[i]}=t_{ij}(x)g_{[j]}. 
\end{equation}
Because $f$ and $\tilde{f}$ satisfy the same gluing condition,
they must be the same object. Thus
we have shown the isomorphism 
\begin{equation}
\Gamma(E_{reg}) \cong C^{\infty}(E_{prin}). 
\end{equation}
In the next subsection, we regard covariant derivatives 
as operators acting on such space.

An isomorphism similar to 
(\ref{isomorphicrelationofVDotimesV}) exists 
for fiber bundles. 
It is given by 
\begin{equation}
\Gamma(E_{r} \otimes E_{reg})  \cong 
\underbrace{ \Gamma(E_{reg}) \oplus \cdots \oplus \Gamma(E_{reg}) 
}_{d_{r}}, 
\label{isobundle}
\end{equation}
where $E_{r}$ is a fiber bundle whose fiber is 
$V_{r}$ and is associated with $E_{reg}$. 
Let $f_{k}^{[i]}(x,g)$ and $f_{(k)}^{[i]}(x,g)$ be 
elements of the left-hand side and right-hand side, 
respectively. 
The isomorphism (\ref{isobundle}) is expressed by the relation 
\begin{equation}
f_{(k)}^{[i]}(x,g)=R_{(k)}{}^{l}(g^{-1})f_{l}^{[i]}(x,g). 
\label{isomorphismonbundle}
\end{equation}
In this way, we can convert the two kinds of indices 
(i.e. indices with and without parentheses) using $R(g)$. 

Using this isomorphism, we can naturally 
lift an element of 
$End(\Gamma(E_{reg}))$ to one of $End(\Gamma(E_{r}\otimes E_{reg}))$. 
This is 
illustrated in Fig. \ref{diagram2}. 
\begin{figure}
$$
\begin{CD}
    \Gamma(E_r\otimes E_{reg})
  @>B>\spadesuit >
  \Gamma(E_r\otimes E_{reg}) \\
  @V{   }V
  {\hspace{2.5cm}\scalebox{1.5}{\large $\circlearrowleft$}}V    
  @V{    }VV\\
  \Gamma(E_{reg})\oplus\cdots\oplus \Gamma(E_{reg})
  @>B\oplus \cdots \oplus B>\spadesuit \spadesuit>
  \Gamma(E_{reg})\oplus\cdots\oplus \Gamma(E_{reg})
\end{CD}
$$
\caption{The action of $B$ commutes with the conversions 
of indices. The latter are expressed by
the vertical arrows and given by the isomorphism 
(\ref{isobundle}). 
} \label{diagram2}
\end{figure} 
The action of $B\in End(\Gamma(E_{reg}))$ 
 on $\Gamma(E_{reg}) \oplus \cdots \oplus \Gamma(E_{reg})$ 
is well-defined, 
because $B$ acts on each component of the direct sum. 
This action is denoted by $\spadesuit \spadesuit$ in 
Fig. $\ref{diagram2}$. 
The action denoted by $\spadesuit$ is defined 
in such a way that the commutative diagram in Fig. 
\ref{diagram2} holds; 
that is, the action $\spadesuit$ is given in terms of 
$\spadesuit \spadesuit$ by 
\begin{equation}
B^{[i]}f_{l}^{[i]}(x,g)=R_{l}{}^{(k)}(g)
B^{[i]}f_{(k)}^{[i]}(x,g). 
\label{generalcommuteAandR}
\end{equation}
Here, $B$ belongs to $End(\Gamma(E_{r} \otimes E_{reg}))$ 
on the left-hand side, 
while it belongs to $End(\Gamma(E_{reg}))$ 
on the right-hand side. 
In this way, we can lift 
$End(\Gamma(E_{reg}))$ to $End(\Gamma(E_{r}\otimes E_{reg}))$. 
Ggeneralizations, 
such as a lift from 
$End(\Gamma(E_{r}\otimes E_{reg}))$ 
to $End(\Gamma(E_{r}\otimes E_{r^{\prime}}\otimes E_{reg})$, 
are straightforward. 
As we see in the next subsecton, 
covariant derivatives are a typical example of this lifting mechanism.

\subsection{Covariant Derivatives as Matrices}
\label{sec:S2example}
\hspace{0.51cm}
As discussed at the beginning of this section, 
we need to prepare a good space 
in which covariant derivatives are expressed as endomorphisms. 
We now show that $\Gamma(E_{reg})$ 
is such space. 
From now on, we take $G$ to be $Spin(10)$ or $Spin^{c}(10)$
\footnote{
The definition of $Spin^{c}(10)$ is given by 
\begin{equation}
Spin^{c}(10)=(Spin(10)\times U(1))/Z_{2}. 
\end{equation}
Every manifold does not necessarily admit a spin structure. 
However, we can introduce a spin-$c$ structure 
into arbitrary manifolds. 
If a manifold we consider admits a spin structure, 
we choose $Spin(10)$. If not, we choose $Spin^{c}(10)$. 
}. 
We consider a ten-dimensional manifold $M$, and take 
a spin structure or a spin-$c$ structure. 

A covariant derivative is given by 
\begin{eqnarray}
\nabla_{a}^{[i]}=e_{a}^{\mu}{}^{[i]}(x)
\left(\partial_{\mu}+\omega_{\mu}^{bc}{}^{[i]}(x)
{\cal O}_{bc} -ia_{\mu}{}^{[i]}(x)\right), 
\end{eqnarray}
where 
${\cal O}_{bc}$ is the generator of $G$ and 
$a_{\mu}(x)$ is a $U(1)$ gauge field belonging to 
the $U(1)$ part of $Spin^{c}(10)$ 
\footnote{
When we consider $Spin(10)$, we can set $a_{\mu}=0$. }. 
When we consider its action 
on $\Gamma(E_{reg})$, 
it can be regarded as the map
\begin{eqnarray}
\nabla_{a}: \Gamma(E_{reg})\rightarrow \Gamma(T \otimes E_{reg}), 
\label{mapofcovariantderivative1}
\end{eqnarray}
where $T$ is the tangent bundle. In general, 
the action of a covariant derivative changes the space. 
Here we use the isomorphism given in 
(\ref{isobundle}),
\begin{eqnarray}
\Gamma(T \otimes E_{reg}) \cong 
\underbrace{\Gamma(E_{reg}) \oplus \cdots \oplus \Gamma(E_{reg})
}_{10}. 
\label{isomorphiccovariantaction}
\end{eqnarray}
By composing the two maps 
(\ref{mapofcovariantderivative1}) and 
(\ref{isomorphiccovariantaction}), 
we obtain the following map:
\begin{eqnarray}
\nabla_{(a)}: \Gamma(E_{reg})\rightarrow 
\underbrace{ \Gamma(E_{reg}) \oplus \cdots \oplus \Gamma(E_{reg})}_{10}.
\label{mapofcovariantderivative2}
\end{eqnarray}
Here, the index $(a)$ is a label indicating one of the ten copies of 
$\Gamma(E_{reg})$. 
Therefore, each component of $\nabla_{(a)}$ can be regarded 
as a mapping from $\Gamma(E_{reg})$ to $\Gamma(E_{reg})$:
\begin{equation}
\nabla_{(a)}\in End(\Gamma(E_{reg}))
\end{equation}
for $a=1,\cdots,10$. 
Explicitly, $\nabla_{(a)}$ and $\nabla_{a}$ 
are related on each patch as 
\begin{equation}
\nabla_{(a)}^{[i]}=R_{(a)}{}^{b}(g_{[i]}^{-1})\nabla_{b}^{[i]}. 
\label{relationbetweentwocovariant}
\end{equation}
Let us confirm the gluing condition for them. 
In the region in which two patches overlap, i.e. 
$x \in U_{i}\cap U_{j}$, 
$\nabla_{a}^{[i]}$ 
and $\nabla_{a}^{[j]}$ 
are related as
\begin{equation}
\nabla_{a}^{[i]}=R_{a}{}^{b}(t_{ij}(x))\nabla_{b}^{[j]}. 
\end{equation}
%$\nabla_{a}$ transforms as a vector under an action of $G$. 
Therfore, we obtain 
\begin{eqnarray}
\nabla_{(a)}^{[i]}&=&R_{(a)}{}^{b}(g_{[i]}^{-1})\nabla_{b}^{[i]}\cr
&=&R_{(a)}{}^{b}(g_{[i]}^{-1})
R_{b}{}^{c}(t_{ij}(x))\nabla_{c}^{[j]}\cr
&=&R_{(a)}{}^{c}((t_{ij}^{-1}g_{[i]})^{-1})\nabla_{c}^{[j]}\cr
&=&\nabla_{(a)}^{[j]}. 
\label{independentofpatch}
\end{eqnarray}
This confirms that 
each component of $\nabla_{(a)}$ is 
globally defined and gives 
an endomorphism. 
In other words, $\nabla_{(a)}$ 
can be expressed as ten
infinite-dimensional matrices. 

We now examine explicitly how each component of $\nabla_{(a)}$ 
acts on $\Gamma(E_{reg})$. 
It acts on an element $f$ of $\Gamma(E_{reg})$ 
as follows: 
\begin{eqnarray}
\nabla_{(a)}^{[i]}f^{[i]}(x,g)
=R_{(a)}{}^{b}(g^{-1})
e_{b}^{\mu}{}^{[i]}(x)
\left(\partial_{\mu}+\omega_{\mu}^{bc}{}^{[i]}(x)
{\cal O}_{bc} -ia_{\mu}{}^{[i]}(x)\right)f^{[i]}(x,g). 
\end{eqnarray}
The action of ${\cal O}_{bc}$ on $\Gamma(E_{reg})$ is given by 
\begin{eqnarray}
i\epsilon^{ab}\left({\cal O}_{ab}f^{[i]}\right)(x,g)
=f^{[i]}\left(x,\left(1+i\epsilon^{ab}M_{ab}\right)^{-1}g\right)
-f^{[i]}(x,g) ,
\label{Oasderivative}
\end{eqnarray}
where $M_{ab}$ is the matrix of the fundamental representation. 
From the relation $\Gamma(E_{reg})\cong C^{\infty}(E_{prin})$, 
we see that this covariant derivative is 
a first-order differential operator on 
$C^{\infty}(E_{prin})$. 

To this point, we have considered a covariant derivative as 
a set of endomorphisms on $\Gamma(E_{reg})$. 
By an argument similar to that 
given in the last part of the previous subsection, 
we can naturally lift 
an endomorphism on $\Gamma(E_{reg})$ to one on 
$\Gamma(E_{r}\otimes E_{reg})$ 
for any representation $r$. 
Let us consider a specific example in which $E_{r}$ is the 
tangent bundle $T$. 
Let $f_{a}^{[i]}(x,g)$ and $f_{(a)}^{[i]}(x,g)$ be elements 
of the left-hand side and the right-hand side 
of (\ref{isomorphiccovariantaction}), respectively. 
Because we know how $\nabla_{(a)}\in {\mbox End}(\Gamma(E_{reg}))$ 
acts on $f_{(b)}^{[i]}(x,g)$, 
we can define 
an endomorphism on $\Gamma(T\otimes E_{reg})$
through the relation 
\begin{equation}
\nabla_{(a)}^{[i]}f_{b}^{[i]}(x,g)=R_{b}{}^{(c)}(g)
\nabla_{(a)}^{[i]}f_{(c)}^{[i]}(x,g). 
\label{liftforcovariantderivative}
\end{equation}
This equation indicates that
\begin{equation}
\epsilon^{de}{\cal O}_{de}f_{b}^{[i]}(x,g)=R_{b}{}^{(c)}(g)
\epsilon^{de}{\cal O}_{de}f_{(c)}^{[i]}(x,g). 
\end{equation}
However, this is merely an infinitesimal version of 
(\ref{indicateGandRcommute}). 
In this sense, the lifting is naturally realized 
for covariant derivatives.

We can now show the hermiticity of $i\nabla_{(a)}$ for each component. 
Let $u$ and $v$ be elements of $\Gamma(E_{reg})$. 
We first define an inner product on $\Gamma(E_{reg})$ as 
\begin{equation}
(u,v)=\int ed^{d}x dg u^{[i]}(x,g)^{\ast}v^{[i]}(x,g), 
\end{equation}
where $e=\det e_{a}^{\mu}$ and $dg$ is the Haar measure. 
This inner product 
is independent of the choice of the patch. 
Then, 
for any $u_{(a)}\in \Gamma(E_{reg})\oplus \cdots \oplus \Gamma(E_{reg})$, 
we can show 
\begin{eqnarray}
(u_{(a)},i\nabla_{(a)}v)
&=&\int ed^{d}xdg u_{(a)}^{[i]}(x,g)^{\ast}
i\nabla_{(a)}^{[i]}v^{[i]}(x,g) \cr
&=&\int ed^{d}x dg u_{a}^{[i]}(x,g)^{\ast}
i\nabla_{a}^{[i]}v^{[i]}(x,g) \cr
&=&\int ed^{d}x dg(i\nabla_{a}^{[i]}u_{a}^{[i]}(x,g))^{\ast}
v^{[i]}(x,g) \cr
&=&\int ed^{d}xdg(i\nabla_{(a)}^{[i]}u_{(a)}^{[i]}(x,g))^{\ast}
v^{[i]}(x,g) \cr 
&=&(i\nabla_{(a)}u_{(a)},v). 
\label{equationhermiticityofnabla}
\end{eqnarray}
To derive this, 
we have used two important relations. The first is 
\begin{eqnarray}
u^{(a)[i]}(x,g)\nabla_{(a)}^{[i]}
%&=&\delta^{(a)(b)}u_{(a)}^{[i]}(x,g)\nabla_{(b)}^{[i]} \cr
&=&\delta^{(a)(b)}R_{(a)}{}^{c}(g^{-1})R_{(b)}{}^{d}(g^{-1})
u_{c}^{[i]}(x,g)\nabla_{d}^{[i]}\cr
&=&u^{c[i]}(x,g)\nabla_{c}^{[i]}, 
\end{eqnarray}
which follows from $R^{T}R=1$, and the second is 
\begin{eqnarray}
\nabla^{(a)[i]}u_{(a)}^{[i]}(x,g)&=&\delta^{(a)(b)}
\nabla_{(a)}^{[i]}u_{(b)}^{[i]}(x,g) \cr
&=&\delta^{(a)(b)}R_{(b)}{}^{d}(g^{-1})
\nabla_{(a)}^{[i]}u_{d}^{[i]}(x,g) \cr
&=&\delta^{(a)(b)}R_{(b)}{}^{d}(g^{-1})R_{(a)}{}^{c}(g^{-1})
\nabla_{c}^{[i]}u_{d}^{[i]}(x,g) \cr
&=&\nabla^{c[i]}u_{c}^{[i]}(x,g) , 
\end{eqnarray}
where we have used the equation 
(\ref{liftforcovariantderivative}) in going 
from the first line to the second line. 
Note that 
$\nabla_{(a)}$ belongs to 
$End(\Gamma(E_{reg}))\oplus \cdots \oplus 
End(\Gamma(E_{reg}))$ in the first line, 
while it belongs to 
$End(\Gamma(T \otimes E_{reg}))$ in the second line. 
Finally, by setting such as 
$u_{(1)}=u$ and $u_{(2)}=\cdots =u_{(10)}=0$ 
in (\ref{equationhermiticityofnabla}), 
we complete the proof of the hermiticity of $i\nabla_{(a)}$.

\vspace{0.3cm}

In the following, 
we present the explicit forms of covariant 
derivatives on $S^{2}$ and $T^{2}$, from which
we can understand how curved spaces are globally 
described by matrices. 

Here we take $G$ to be $Spin(2)$. 
Its representation of spin $s$ is given by 
\begin{eqnarray}
  R^{ \langle s \rangle}(\theta)=e^{2is\theta}
  \quad
  (\theta\in[0,2\pi)). 
  \label{U(1)repmatrix}
\end{eqnarray}
As shown in (\ref{Oasderivative}), 
the Lorentz generator ${\cal O}_{+-}$ is expressed 
in terms of  
the derivative with respect to $\theta$ as
\begin{eqnarray}
  {\cal O}_{+-}=\frac{1}{4i}\frac{\partial}{\partial\theta}. 
\end{eqnarray}
Here, $+$ and $-$
indicate the linear combinations of the Lorentz
indices $1+i2$ and $1-i2$, respectively.

First we consider $S^{2}$ with a homogeneous and isotropic metric.
In the stereographic coordinates 
projected from the north pole, we have 
\begin{eqnarray}
  g_{z\bar{z}}=g_{\bar{z}z}=\frac{1}{(1+z\bar{z})^2} , 
  \quad
  g_{zz}=g_{\bar{z}\bar{z}}=0 . 
  \end{eqnarray}
The simplest choice of the zweibein consistent with this 
metric is 
\begin{eqnarray}
  e_{+\bar{z}}=e_{-z}=\frac{1}{1+z\bar{z}} , 
  \quad
  e_{+z}=e_{-\bar{z}}=0,  
\end{eqnarray}
and the spin connection is given by
\begin{eqnarray}
  \omega_{z}^{+-}=-\frac{\bar{z}}{1+z\bar{z}} , 
  \quad
  \omega_{\bar{z}}^{+-}=\frac{z}{1+z\bar{z}} . 
\end{eqnarray}
Following the general procedure 
(\ref{relationbetweentwocovariant}),
we obtain
\begin{eqnarray}
  &&\nabla_{(+)}^{[z]}
  =
  e^{-2i\theta}
  \left(
    (1+z\bar{z})\partial_z
    +
    \frac{i}{2}\bar{z}\partial_\theta 
  \right) , 
  \cr
  &&\nabla_{(-)}^{[z]}
  =
  e^{2i\theta}
  \left(
    (1+z\bar{z})\partial_{\bar{z}}
    -
    \frac{i}{2}z\partial_\theta
  \right) .  
  \label{matrixcovariantS2}
\end{eqnarray}
 
We can explicitly check that these 
satisfy the gluing condition given in (\ref{independentofpatch}). 
We introduce stereographic coordinate 
$(w,\bar{w})$ projected from the south pole. 
In the region in which the two patches overlap, 
they are related by $z=1/w$.  
The coordinate  $\theta^{\prime}$ of the fiber on this patch 
is related to $\theta$ by the transition function
\begin{eqnarray}
  \theta^\prime
  =\theta+{{\rm arg}}(z)+\frac{\pi}{2} .   
\end{eqnarray}
Rewriting (\ref{matrixcovariantS2}) in terms of 
$(w,\theta^{\prime})$, we obtain 
\begin{eqnarray}
  &&\nabla_{(+)}^{[w]}
  =
  e^{-2i\theta^\prime}
  \left(
    (1+w\bar{w})\partial_w
    +
    \frac{i}{2}\bar{w}\partial_{\theta^\prime} 
  \right), \cr
   && \nabla_{(-)}^{[w]}
  =
  e^{2i\theta^\prime}
  \left(
    (1+w\bar{w})\partial_{\bar{w}}
    -\frac{i}{2}w\partial_{\theta^\prime} 
  \right), 
\end{eqnarray}
which have the same forms as (\ref{matrixcovariantS2}) 
as expected. 
This confirms that 
the index $(a)$ does not transform under a local Lorentz 
transformation, and each component of $\nabla_{(a)}$ 
is a scalar operator.
Thus, a two-sphere can be expressed in terms of 
two endomorphisms. 

Because $\nabla_{(+)}$ and $\nabla_{(-)}$ are endomorphisms, 
their products are well-defined,
and we can consider their commutators.
A simple calculation gives 
\begin{eqnarray}
  [\nabla_{(+)},\nabla_{(-)}]
  =
  -i\frac{\partial}{\partial\theta}
  =
  4 {\cal O}_{+-}
\end{eqnarray}
and 
\begin{eqnarray}
  [{\cal O}_{+-},\nabla_{(\pm)}]
  &=&
  \mp \frac{1}{2}\nabla_{(\pm)}. 
\end{eqnarray}
Thus 
$\nabla_{(\pm)}$ and ${\cal O}_{+-}$ form the 
$\mathfrak{su}(2)$ algebra. 
Therefore we have
\begin{equation}
[\nabla_{(b)},[\nabla^{(b)},\nabla_{(a)}]]=-\nabla_{(a)}, 
\end{equation}
for $a=1,2$ ,
and it turns out that the homogeneous isotropic two-sphere 
is a classical solution of a two-dimensional matrix model 
with a negative mass term (see also 
section \ref{sec:classicalsolutions}). 

Here we emphasize the difference from a fuzzy two-sphere. 
It is described 
by an embedding into the three-dimensional flat 
space, and we identify the $SO(3)$ symmetry of this 
space with the isometry of that sphere. 
We thus need three matrices, i.e. 
derivatives are given by angular momentum operators. 
It is not described in terms of two matrices, 
although it is a two-dimensional manifold.

\vspace{0.3cm}

\begin{figure}[htb]
  \begin{center}
    \scalebox{0.40}{
      \includegraphics[25cm,16cm]{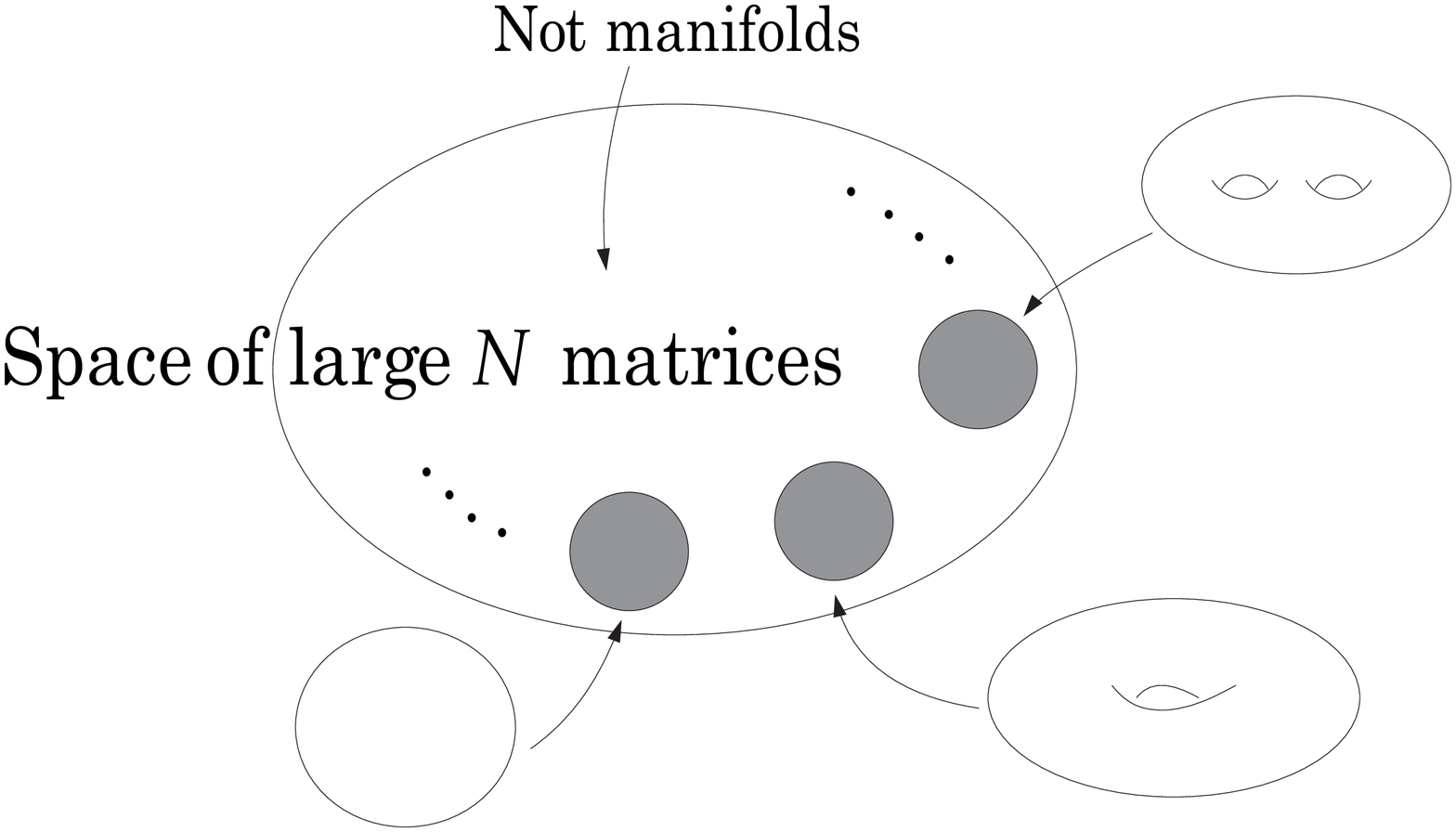}
    }
    \caption{
    All $d$-dimensional manifolds 
    can be described by 
    $d$ large $N$ matrices. }
    \label{figure}
  \end{center}
\end{figure}

We next consider a second example, $T^{2}$. 
To cover all regions of $T^{2}$, we need four coordinate patches. 
We denote them by $(z_{i},\bar{z}_{i})$ $(i=1,\cdots,4)$. 
The simplest metric in this case is 
\begin{eqnarray}
  g_{z_{i}\bar{z_{i}}}=g_{\bar{z_{i}}z_{i}}=1 , 
  \quad
  g_{z_{i}z_{i}}=g_{\bar{z_{i}}\bar{z_{i}}}=0 , 
\end{eqnarray}
and the corresponding covariant derivatives are given by
\begin{eqnarray}
  &&\nabla_{(+)}^{[z_{i}]}
  =
  e^{-2i\theta_{i}}\partial_{z_{i}}, 
  \cr
  &&\nabla_{(-)}^{[z_{i}]}
  =e^{2i\theta_{i}}\partial_{\bar{z_{i}}}.
\end{eqnarray}
In a region in which coordinate patches overlap, 
they are related by 
$z_{i}=z_{j}+c_{ij}$, where $c_{ij}$ is a constant. 
The fiber is trivially glued as 
\begin{eqnarray}
\theta_{i}=\theta_{j}. 
\end{eqnarray}
We have seen that two-dimensional manifolds 
with different topologies can be uniformly described by 
two infinite-dimensional matrices. 
In a similar way, all $d$-dimensional manifolds can be 
described by $d$ infinite-dimensional matrices. 
(See Fig. \ref{figure}.)

%%%%%%%%%%%%%%%%%%%%%%%%%%%%%%%%%%%%%%%%%%%%%%%%%%%%%%%%%%%%%%%%%%%%%%%%%%%%%%

\section{New Interpretation of Matrix Model}
\label{sec:matrixmodelandloallorentz}
\hspace{0.51cm}
In this section, we apply the idea developed in 
the previous section to matrix models. 
We have seen that covariant derivatives
can be expressed as endomorphisms 
if their indices are converted
from $a$ to $(a)$. 
For this reason, 
it is natural to regard the matrix model (\ref{IIBaction}) 
to be written in terms of indices with parentheses: 
\begin{eqnarray}
 && S=-\frac{1}{4g^{2}}tr\left([A_{(a)},A_{(b)}]
    [A_{(c)},A_{(d)}]\delta^{(a)(c)}\delta^{(b)(d)}\right) \cr
  &&\hspace{1.0cm} 
    +\frac{1}{2g^{2}}tr\left(
  \bar{\psi}^{(\alpha)}(\Gamma^{(a)})_{(\alpha)}^{\hspace{0.2cm}(\beta)}
  [A_{(a)},\psi_{(\beta)}]\right).  
  \label{matrixmodelwith()indices}  
\end{eqnarray}
Each component of the dynamical variables can be identified with
an endomorphism as 
\begin{eqnarray}
A_{(a)}: \Gamma(E_{reg})\rightarrow 
\Gamma(E_{reg}),
\end{eqnarray}
where $(a)=1,\cdots 10$, and 
\begin{eqnarray}
\psi_{(\alpha)}: \Gamma(E_{reg})\rightarrow 
\Gamma(E_{reg}),
\end{eqnarray}
where $(\alpha)=1,\cdots 16$. 
According to the equation (\ref{isomorphismonbundle}), 
we can convert the indices as 
\begin{equation}
A_{a}=R_{a}{}^{(b)}(g)A_{(b)}, 
%A_{(a)}=R_{(a)}{}^{b}(g^{-1})A_{b}, 
\end{equation}
where $R_{a}{}^{(b)}(g)$ is the vector representation, and 
we have 
\begin{equation}
\psi_{\alpha}=R_{\alpha}{}^{(\beta)}(g)\psi_{(\beta)}, 
\end{equation}
where $R_{\alpha}{}^{(\beta)}(g)$ is the spinor representation. 
In particular, 
$A_{a}$ provides a mapping from $\Gamma(E_{reg})$ 
to $\Gamma(T \otimes E_{reg})$, where $T$ is the tangent bundle, 
and $\psi_{\alpha}$ provides a mapping from 
$\Gamma(E_{reg})$ 
to $\Gamma(S \otimes E_{reg})$, where $S$ is a spin bundle. 
Note that $A_{(a)}$ and $\psi_{(\alpha)}$ can be expressed 
as matrices, while $A_{a}$ and $\psi_{\alpha}$ cannot. 

We next attempt to rewrite the above 
action in terms of $A_{a}$ and $\psi_{\alpha}$. 
We first point out the following fact. 
In a quantity that is constructed from the product of 
matrices $A_{(a)}$, 
the indices with parentheses can be converted to 
those without parentheses by 
multiplying $R(g)$ on the left: 
\begin{equation}
A_{(a)}A_{(b)}A_{(c)}\cdots
=R_{(a)}{}^{a^{\prime}}(g^{-1})R_{(b)}{}^{b^{\prime}}(g^{-1})
R_{(c)}{}^{c^{\prime}}(g^{-1})
A_{a^{\prime}}A_{b^{\prime}}A_{c^{\prime}}\cdots. 
\label{productinterchangeindices}
\end{equation}
Here, the right-hand side is defined
by use of the lifting discussed in 
the last part of subsection \ref{sec:preliminaries}. 
For an endomorphism $A_{(a)}\in End(\Gamma(E_{reg}))$, 
we can naturally lift it to 
$A_{(a)}\in End(\Gamma(E_{r}\otimes \cdots \otimes E_{reg}))$ 
in such a way that the action of $A_{(a)}$ commutes 
with the conversion of indices. 
For example, $A_{(a)}\in End(\Gamma(T \otimes E_{reg}))$ 
is defined in such a way that the following equation is satisfied: 
\begin{equation}
A_{(a)}f_{(b)}(x,g)=R_{(b)}{}^{c}(g^{-1})A_{(a)}f_{c}(x,g). 
\end{equation}
Here, $A_{(a)}$ belongs to $End(\Gamma(E_{reg}))$ 
on the left-hand side, 
while it belongs to $End(\Gamma(T\otimes E_{reg}))$ 
on the right-hand side. 
From this, we can show
\begin{eqnarray}
A_{(a)}A_{(b)}f(x,g)
&=&A_{(a)}R_{(b)}{}^{b^{\prime}}(g^{-1})A_{b^{\prime}}f(x,g) \cr
&=&R_{(b)}{}^{b^{\prime}}(g^{-1})A_{(a)}A_{b^{\prime}}f(x,g) \cr
&=&R_{(a)}{}^{a^{\prime}}(g^{-1})R_{(b)}{}^{b^{\prime}}(g^{-1})
A_{a^{\prime}}A_{b^{\prime}}f(x,g) . 
\end{eqnarray}
We stress that the product $A_{a^{\prime}}A_{b^{\prime}}$ 
is not the conventional matrix product. 
Rather, 
$A_{a^{\prime}}$ acts on the index $b^{\prime}$ 
as does an ordinary covariant derivative. 
In a similar manner, we can prove 
the equation (\ref{productinterchangeindices}), 
which can be straightforwardly 
extended to include 
the fermionic matrices $\psi_{(\alpha)}$. 

When indices are contracted on 
the left-hand side of $(\ref{productinterchangeindices})$, 
some of the representation matrices $R$ 
on the right-hand side cancel out, due to the relation $R^{T}R=1$.  
For instance, we obtain 
\begin{eqnarray}
\delta^{(a)(b)}A_{(a)}A_{(b)}=\delta^{ab}A_{a}A_{b}
\label{contracted}
\end{eqnarray}
and 
\begin{eqnarray}
\bar{\psi}^{(\alpha)}(\Gamma^{(a)})_{(\alpha)}{}^{(\beta)}
A_{(a)}\psi_{(\beta)}
&=&
R^{(\alpha)}{}_{\gamma}(\Gamma^{(a)})_{(\alpha)}{}^{(\beta)}
R_{(a)}{}^{b}R_{(\beta)}{}^{\delta}
\bar{\psi}^{\gamma}A_{b}\psi_{\delta} \cr
&=&\bar{\psi}^{\gamma}(\Gamma^{b})_{\gamma}{}^{\delta}
A_{b}\psi_{\delta}. 
\end{eqnarray}
Thus, 
the above matrix model action can be brought to the following form: 
\begin{eqnarray}
  S=-\frac{1}{4g^{2}}tr\left([A_{a},A_{b}]
    [A^{c},A^{d}]\delta^{ac}\delta^{bd}\right) 
    +\frac{1}{2g^{2}}tr\left(
  \bar{\psi}^{\alpha}(\Gamma^{a})_{\alpha}^{\hspace{0.2cm}\beta}
  [A_{a},\psi_{\beta}]\right).
  \label{matrixmodel2}
\end{eqnarray}
We stress that 
$A_{a}$ and $\psi_{\alpha}$ are 
no longer matrices. 
Their product is taken like that of 
ordinary covariant derivatives.

%%%%%%%%%%%%%%%%%%%%%%%%%%%%%%%%%%%%%%%%%%%%%%%%%%%%%%%%%%%%%%%%%%%%%%%%%%%%%%%

\subsection{Classical Solutions}
\label{sec:classicalsolutions}
\hspace{0.51cm}
In this subsection, we study classical solutions. 
The equation of motion is given by 
\begin{equation}
[A^{(a)},[A_{(a)},A_{(b)}]]=0 
\end{equation}
in the case $\psi=0$. 
Taking account of (\ref{productinterchangeindices}) and 
(\ref{contracted}), we find that 
this is equivalent to 
\begin{equation}
[A^{a},[A_{a},A_{b}]]=0. 
\label{equationofmotion}
\end{equation}
We search for classical solutions having the following form: 
\begin{equation}
A_{a}=i\nabla_{a}(x)=ie_{a}^{\mu}(x)
(\partial_{\mu}+\omega_{\mu}^{bc}(x){\cal O}_{bc}-ia_{\mu}(x)).
\label{ansatzforclassicalsolution}
\end{equation}
Here we assume that 
there is no torsion, for simplicity. 
Then, 
substituting this into the left-hand side of (\ref{equationofmotion}), 
we obtain
\footnote{Note that 
$[{\cal O}_{cd},\nabla_{a}]=\frac{1}{2}
(\eta_{ca}\nabla_{d}-\eta_{da}\nabla_{c})$ and 
$R_{ab}{}^{ac}=R_{b}{}^{c}$. } 
\begin{eqnarray}
[\nabla^{a},[\nabla_{a},\nabla_{b}]]
&=&[\nabla^{a},R_{ab}{}^{cd}{\cal O}_{cd}-if_{ab}] \cr
&=&[\nabla^{a},R_{ab}{}^{cd}]{\cal O}_{cd}
+R_{ab}{}^{cd}[\nabla^{a},{\cal O}_{cd}]
-i[\nabla^{a},f_{ab}] \cr
&=&
(\nabla^{a}R_{ab}{}^{cd}){\cal O}_{cd}-R_{b}{}^{c}\nabla_{c}
-i(\nabla^{a}f_{ab}). 
\label{calculation}
\end{eqnarray}
Therefore, (\ref{equationofmotion}) is satisfied if 
the following hold:
\begin{eqnarray}
\nabla^{a}R_{ab}{}^{cd}=0, \quad R_{bc}=0, \quad 
\nabla^{a}f_{ab}=0. 
\label{equationofmotionRF}
\end{eqnarray}
The first equation follows from 
the second by 
the Bianchi identity 
$\nabla_{[a}R^{bc}{}_{de]}=0$.
Thus we have derived the Ricci-flat condition 
and the Maxwell equation 
from the equation of motion of the matrix model.

The above result seems unnatural in the sense that
the energy-momentum tensor of $a_{\mu}$
does not appear in the second equation. 
It might be possible to resolve with the following argument.
If we expand the exact equations
with respect to the string scale $\alpha^{\prime}$,
they would take forms like the following:
\begin{eqnarray}
&&R_{ab}-\frac{1}{2}\delta_{ab}R\sim 
\alpha^{\prime}\left( f_{a}{}^{c}f_{cb}
+\frac{1}{4}\delta_{ab}f^{2}\right), \cr
&&\nabla^{a}f_{ab}\sim O(\alpha^{\prime}).
\end{eqnarray}
We thus see that 
the equations we have obtained can
be regarded as the lowest-order forms of the exact equations. 
If the loop expansion of the matrix model 
is in some manner related to the $\alpha^{\prime}$ expansion, 
the classical solution naturally has
the form of that which we have obtained.
And the contribution from the energy momentum ternsor 
comes from quantum corrections 
\footnote{
This is consistent with the 
original interpretation of the matrix model \cite{BFSS,IKKT}, 
where the coupling between 
the gravity and the energy momentum tensor 
arises as a one-loop quantum effect.}.
This result might suggest a closer relation between
our new interpretation of the matrix model
and the world-sheet picture of string theory,
because in the latter, the loop expansion is 
nothing but the $\alpha^{\prime}$ expansion.

We next consider a matrix model with a mass term, 
\begin{eqnarray}
  S=-\frac{1}{4g^{2}}tr\left([A_{(a)},A_{(b)}]
    [A^{(a)},A^{(b)}]\right) +\frac{m^{2}}{g^{2}}
    tr\left(A_{(a)}A^{(a)}\right). 
\end{eqnarray}
The equation of motion is now given by 
\begin{equation}
[A_{(b)},[A_{(a)},A^{(b)}]]-2m^{2}A_{(a)}=0, 
\end{equation}
and a calculation similar to (\ref{calculation}) gives
\begin{equation}
\nabla^{a}R_{ab}{}^{cd}=0, \quad R_{bc}=-2m^{2}\delta_{bc}
\quad \nabla^{a}f_{ab}=0. 
\end{equation}
The first condition again follows from 
the Bianchi identity and the second equation. 
The second equation gives the Einstein equation 
with the cosmological constant 
\begin{equation}
\Lambda=-(d-2)m^{2}. 
\end{equation}
This implies that maximally symmetric spaces are classical solutions. 
A positive (negative) mass term 
corresponds to 
a negative (positive) cosmological constant.

%%%%%%%%%%%%%%%%%%%%%%%%%%%%%%%%%%%%%%%%%%%%%%%%%%%%%%%%%%%%%%%%%%%%%%%%%%%%%%

\subsection{Expansion with respect to Covariant Derivatives}
\label{sec:newmatrixspace}
\hspace{0.51cm}
As we have seen, each component of the matrices 
$A_{(a)}$ and $\psi_{(\alpha)}$ 
can be regarded as 
an element of 
$End(C^{\infty}(E_{prin}))$. 
Therefore it can be expanded 
by derivatives acting on $E_{prin}$, that is, 
the covariant derivatives $\nabla_{a}$ 
and the Lorentz generators ${\cal O}_{ab}$ 
\footnote{
If $G$ is $Spin^{c}(10)$, we have a $U(1)$ generator 
in addition to them. Here we ignore it for simplicity. 
}. 
In genral, the coefficients in such an expansion 
are functions on $C^{\infty}(E_{prin})$, which 
are expanded as 
\begin{equation}
f^{[i]}(x,g)=\sum_{r:rep}c_{ij}^{[i]\langle r \rangle}(x)
R_{ij}^{\langle r \rangle}(g),
\label{coefficientexpansiononbundle}
\end{equation}
where the summation is taken over all irreducible representations. 
For details, see 
Appendix \ref{sec:proofofregulardecomposition}. 
As there seem to be too many fields,
it would be better if we could 
restrict this space to a smaller one.
We show here that 
it is indeed possible to remove the $g$-dependence in 
(\ref{coefficientexpansiononbundle}) by imposing a physically 
natural constraint. 

First, we introduce the right action of $G$ on $C^{\infty}(E_{prin})$,
and denote it 
by $\hat{r}(h)$: 
\begin{equation}
\hat{r}(h) : f^{[i]}(x,g) \mapsto f^{[i]}(x,gh). 
\end{equation}
Note that this right action commutes with the left one. 
For $A_{(a)}$, 
we require that it transforms as a vector 
under the right action of $G$: 
\begin{eqnarray}
\hat{r}(h^{-1})A_{(a)}\hat{r}(h)
&=&R_{(a)}{}^{(b)}(h)A_{(b)}. 
\label{requirementforboson()}
\end{eqnarray}
Because the left-hand side can be rewritten as 
\begin{eqnarray}
\hat{r}(h^{-1})A_{(a)}\hat{r}(h)
&=&\hat{r}(h^{-1})R_{(a)}{}^{b}(g^{-1})A_{b}\hat{r}(h) \cr
&=&R_{(a)}{}^{b}(hg^{-1})\hat{r}(h^{-1})A_{b}\hat{r}(h) , 
\end{eqnarray}
the requirement (\ref{requirementforboson()})
is equivalent to 
\begin{eqnarray}
\hat{r}(h^{-1})A_{a}\hat{r}(h)=A_{a}. 
\label{requirementforboson}
\end{eqnarray}
If $A_{a}$ is a function given by $A_{a}=a_{a}^{[i]}(x,g)$, 
the above equation imposes the condition 
$a_{a}^{[i]}(x,g)=a_{a}^{[i]}(x,gh^{-1})$, which means 
that $a_{a}^{[i]}$ is independent of $g$. 
The covariant derivative $\nabla_{a}$ also satisfies the constraint 
(\ref{requirementforboson}), because 
${\cal O}_{ab}$ commutes with the right action. 
Therefore, 
the expansion of $A_{a}$ is given by
\footnote{
We can assume that these coefficients are symmetric under 
permutations of 
$\nabla_{a}$ and ${\cal O}_{bc}$. 
For example, we have 
$a_{(2)}{}_{a}^{bc,de}=a_{(2)}{}_{a}^{de,bc}$ and 
$a_{a}^{bc}=a_{a}^{cb}$. 
Their anti-symmetric parts can be absorbed into 
$a_{(1)}{}_{a}^{bc}$, because 
we have 
$a_{(2)}{}_{a}^{[bc,de]}{\cal O}_{bc}{\cal O}_{de}
=-a_{(2)}{}_{a}^{[ce,cb]}{\cal O}_{eb}$ and 
$a_{a}^{[bc]}\nabla_{b}\nabla_{c}=
\frac{1}{2}a_{a}^{[bc]} R_{bc}{}^{de}{\cal O}_{de}$. 
}
\begin{eqnarray}
  A_{a}
  =\hat{a}_{a}^{[i]}(x)+
  \frac{i}{2}\{\hat{a}_{a}^{b[i]}(x), \nabla_b^{[i]} \}
  +\frac{i^{2}}{2}\{\hat{a}_{a}^{bc[i]}(x), 
  \nabla_{b}^{[i]} \nabla_{c}^{[i]}\}
  +\cdots, 
  \label{expansionA-x}
\end{eqnarray}
where
\begin{eqnarray}
 &&\hat{a}_{a}^{[i]}(x) =a_{a}^{[i]}(x)+
 \frac{i}{2}
  \left\{
    a_{(1)}{}_{a}^{bc[i]}(x),
    {\cal O}_{bc}\right\}
  +
  \frac{i^{2}}{2}
  \left\{
    a_{(2)}{}_{a}^{bc,de[i]}(x),
    {\cal O}_{bc}{\cal O}_{de}
  \right\}
  +\cdots,  \cr
 && \hat{a}_{a}^{b\cdots[i]}(x)
  =
  a_{a}^{b\cdots[i]}(x)+
  \frac{i}{2}
  \left\{
    a_{(1)}{}_{a}^{b\cdots, cd[i]}(x),
    {\cal O}_{cd}
  \right\} \cr
 && \hspace{2.0cm}+
  \frac{i^{2}}{2}
  \left\{
    a_{(2)}{}_{a}^{b\cdots ,cd,de[i]}(x),
    {\cal O}_{cd}
    {\cal O}_{de}
  \right\}
  +\cdots. 
  \label{bosoniccoefficientexpansionbyO}
\end{eqnarray}
Here, the hat symbol $\hat{}\ $ 
indicates a power series in ${\cal O}_{ab}$. 
Here, the anti-commutators have been introduced 
to make each term manifestly 
hermitian. 

We require that $\psi_{(\alpha)}$ 
transforms as a spinor 
under the right action of $G$: 
\begin{equation}
\hat{r}(h^{-1})\psi_{(\alpha)}\hat{r}(h)
=R_{(\alpha)}{}^{(\beta)}(h)\psi_{(\beta)}. 
\end{equation}
This is again equivalent to the requirement that 
$\psi_{\alpha}$ transforms as a scalar, 
\begin{equation}
\hat{r}(h^{-1})\psi_{\alpha}\hat{r}(h)
=\psi_{\alpha}, 
\end{equation}
and therefore $\psi_{\alpha}$ can be expanded similarly 
to (\ref{expansionA-x}). 
 
Finally, we require that 
the infinitesimal parameter 
of the unitary transformation $\Lambda$ 
transforms as a scalar 
under the right action of $G$: 
\begin{equation}
\hat{r}(h^{-1})\Lambda \hat{r}(h)
=\Lambda. 
\end{equation}
Therefore $\Lambda$ can also be expanded similarly to (\ref{expansionA-x}). 

The transformation law 
under the right action of $G$ is summarized as follows: 
$\Lambda$, 
$A_{(a)}$ and $\psi_{(\alpha)}$ 
transform as a scalar, vector and spinor, respectively, 
while 
$A_{a}$ and $\psi_{\alpha}$ transform as scalars. 

\vspace{0.3cm}

Next, we confirm the background independence of
the expansion (\ref{expansionA-x}).
If we ignore the hermiticity for simplicity, 
the expansion can be expressed as 
\footnote{
For the remainder of this paper, we suppress the symbol $[i]$. 
}
\begin{eqnarray}
\hat{a}_{a}(x)+\hat{a}_{a}^{b}(x)\nabla_{b}
+\hat{a}_{a}^{bc}(x)\nabla_{b}\nabla_{c}+\cdots. 
\label{bgdependform}
\end{eqnarray} 
Let us see what happens if we 
express this using 
a different background $e_{a}^{\prime \mu}(x)$ 
and $\omega_{\mu}^{\prime ab}(x)$. 
Then, a simple calculation shows 
\begin{eqnarray}
\hat{a}_{a}^{b}(x)\nabla_{b}
&=&\hat{a}_{a}^{b}(x)e_{b}^{\mu}(x)\left(\partial_{\mu}
+\omega_{\mu}^{cd}(x){\cal O}_{cd}\right)\cr
&=&\hat{a}_{a}^{\mu}(x)\left(\partial_{\mu}
+\omega_{\mu}^{cd}(x){\cal O}_{cd}\right)\cr
&=&\hat{a}_{a}^{\mu}(x)\left(\partial_{\mu}
+\omega_{\mu}^{\prime cd}(x){\cal O}_{cd}\right)
+\hat{a}_{a}^{\mu}(x)
\delta \omega_{\mu}^{cd}(x){\cal O}_{cd} \cr
&=&\hat{a}_{a}^{\mu}(x)\nabla_{\mu}^{\prime}
+\hat{a}_{a}^{\mu}(x)
\delta \omega_{\mu}^{cd}(x){\cal O}_{cd} \cr
&=&\hat{a}_{a}^{\prime b }(x)\nabla_{b}^{\prime}
+\hat{a}_{a}^{\prime b }(x)e_{b}^{\prime \mu}(x)
\delta \omega_{\mu}^{cd}(x){\cal O}_{cd} . 
\label{changebackground}
\end{eqnarray}
where 
$\delta \omega_{\mu}^{cd}(x)=
\omega_{\mu}^{cd}(x)-\omega_{\mu}^{\prime cd}(x)$. 
Although the second term 
in the last line of (\ref{changebackground}) 
seems to be an extra one,  
taking account of the fact that $\hat{a}_{a}(x)$ is given by 
a power series in ${\cal O}_{ab}$ as 
\begin{equation}
\hat{a}_{a}(x)=a_{a}(x)+a_{a}^{bc}(x){\cal O}_{bc}
+a_{a}^{bc,de}(x){\cal O}_{bc}{\cal O}_{de}
+\cdots, 
\end{equation}
we find that 
it can be absorbed into $a_{a}^{bc}(x)$. 
Through similar calculations, 
the expansion (\ref{bgdependform}) can be 
brought into the same form, after we change the background: 
\begin{eqnarray}
\hat{a}_{a}^{\prime}(x)+\hat{a}_{a}^{\prime b}(x)\nabla_{c}^{\prime}
+\hat{a}_{a}^{\prime bc}(x)
\nabla_{b}^{\prime}\nabla_{c}^{\prime}+\cdots.
\end{eqnarray}
This demonstrates the background independence of 
(\ref{bgdependform}). 

\vspace{0.4cm}

Finally, we comment on supersymmetry. 
The matrix model (\ref{matrixmodelwith()indices}) 
is invariant under the following supersymmetry transformation, 
\begin{eqnarray}
  &&\delta^{(1)} A_{(a)}=
  i\bar{\epsilon}^{(\alpha)}
  (\Gamma_{(a)})_{(\alpha)}{}^{(\beta)}
  \psi_{(\beta)}, \cr
  &&\delta^{(1)} \psi_{(\alpha)}
  =\frac{i}{2}[A_{(a)},A_{(b)}]
  (\Gamma^{(a)(b)})_{(\alpha)}{}^{(\beta)}\epsilon_{(\beta)}, 
\end{eqnarray}
where $\epsilon^{(\alpha)}$ is a constant parameter. 
We have restricted the matrix space by 
requiring that 
$A_{(a)}$ and $\psi_{(\alpha)}$ transform 
as a vector and a spinor, respectively, 
under the right action of $G$. 
Accordingly, we have to require that 
$\epsilon_{(\alpha)}$ transforms as a spinor. 
There is, however, no such constant parameter. 
Therefore, 
the restriction under the right action of $G$ 
destroys the supersymmetry. 
We discuss this point further in section 
\ref{sec:conclusiondiscussion}.

\subsection{Diffeomorphism and Local Lorentz Symmetry}
\label{sec:extendedmatrixmodel}
\hspace{0.51cm}
In this subsection, we analyze the unitary symmetry of the matrix model. 
In particular, we show that the differmorphism and the local Lorentz 
symmetry are included in it. 
The matrix model (\ref{matrixmodelwith()indices}) 
has the unitary symmetry 
\begin{eqnarray}
  \delta A_{(a)} = i[ \Lambda,A_{(a)} ] , 
  \quad \delta \psi_{(\alpha)} = i[\Lambda,\psi_{(\alpha)} ],  
  \label{unitarytransformation()}
\end{eqnarray}
or, equivalently, 
\begin{eqnarray}
  \delta A_{a} = i[ \Lambda,A_{a} ] , 
  \quad \delta \psi_{\alpha} = i[\Lambda,\psi_{\alpha} ].    
  \label{unitarytransformation}
\end{eqnarray}

We first consider the $U(1)$ gauge transformation, 
which is generated by $\Lambda=\lambda(x)$.
Because $a_{a}$ transforms according to 
\begin{eqnarray}
 && \delta a_{a} (x)=\partial_{a}\lambda(x),
 \label{U(1)gaugesymm}
\end{eqnarray}
we can identify $a_{a}$ with the corresponding gauge field.

The diffeomorpshim is generated by 
$\Lambda=
\frac{i}{2}\{\lambda^{\mu}(x),\partial_{\mu}\}$.
For example, the fields that appear in the first terms in 
the expansion (\ref{bosoniccoefficientexpansionbyO}) 
transform according to 
\begin{eqnarray}
&&\delta a_{a}(x)
%=i[i\lambda^{\mu}(x)\partial_{\mu},a_{a}(x)]
=-\lambda^{\mu}(x)\partial_{\mu}a_{a}(x), 
\label{diffeotra_{a}}\\
&&\delta a_{a}^{\mu}(x)
%=i[i\lambda^{\mu}\partial_{\mu},a_{a}^{\nu}\partial_{\nu}]
=-\lambda^{\nu}(x)\partial_{\nu}a_{a}^{\mu}(x)
+a_{a}^{\nu}(x)\partial_{\nu}\lambda^{\mu}(x), 
\label{diffvielbein}
\\
&&\delta a_{a}^{\mu\nu}(x)
=-\lambda^{\rho}(x)\partial_{\rho}a_{a}^{\mu\nu}(x)
+2a_{a}^{\rho(\mu}(x)\partial_{\rho}\lambda^{\nu)}(x), 
\label{diffforspinthree}
\end{eqnarray}
where the parentheses $($ $)$ represent the symmetrization 
operation 
\footnote{Note that 
$a^{(\mu_{1}\cdots \mu_{n})}\equiv(a^{\mu_{1}\dots\mu_{n}}+
\mbox{all permutations of the indices})/n!$. 
}. 
In general, the field $a_{a}^{\mu_{1}\cdots\mu_{s-1}}(x)$ 
transforms as a rank-$(s-1)$ symmetric tensor field. 

\vspace{0.4cm} 
Finally, we discuss the local Lorentz symmetry, which is obtained by
taking $\Lambda=\lambda_{(1)}{}^{ab}(x){\cal O}_{ab}$.
Here, the subscript $(1)$ is added to indicate 
the first power of ${\cal O}_{ab}$. 
The background vielbein $e_{a}^{\mu}(x)$ and 
the spin connection $\omega_{\mu}^{ab}(x)$ indeed transform 
in accordance with 
\begin{eqnarray}
 && 
 \delta e_a^\mu (x)=
  -\lambda_{(1)a}{}^{b}(x)e_{b}^{\mu}(x) ,
  \label{LLforbackgrondvielbein} \\
 &&\delta
  \omega_\mu{}^{ab}(x) = \partial_\mu \lambda_{(1)}{}^{ab}(x) 
  +2\lambda_{(1)}{}^{[a}{}_{c}(x)\omega_{\mu}^{b]c}(x), 
  \label{LLforspinconnection}
 \end{eqnarray}
which are obtained from 
$\delta \nabla_{a}=-[\lambda_{(1)}{}^{bc}(x){\cal O}_{bc},\nabla_{a}]$. 
The transformations for the gauge field $a_{a}(x)$, 
the vielbein field $a_{a}^{b}(x)$ and 
the fluctuation of the spin connection $a_{(1)}{}_{a}^{bc}(x)$ 
are also obtained as 
\begin{eqnarray}
  \delta a_a (x)
  &=&
  -\lambda_{(1)a}{}^{b}(x)a_b (x), 
  \\
    \delta a_a{}^{b}(x)
  &=&
  -\lambda_{(1)a}{}^{c} (x)a_{c}{}^{b}(x)
  -\lambda_{(1)}{}^{b}{}_{c}(x) a_{a}{}^{c}(x),\\
    \delta
  a_{(1)}{}_{a}^{bc}(x) &=& -\lambda_{(1)}{}_{a}{}^{d}(x)
  a_{(1)}{}_{d}^{bc}(x)
  +2\lambda_{(1)}^{[b}{}_{d}(x)a_{(1)}{}_{a}^{c]d}(x). 
\end{eqnarray}
For the fermionic field $\chi(x)$, we obtain
\begin{eqnarray}
  \delta \chi(x)&=& 
  -\frac{1}{4}\lambda_{(1)}{}^{bc}(x)\Gamma_{bc}\chi(x).
\end{eqnarray}
In this way, 
the local Lorentz transformation is correctly reproduced. 

Other gauge symmetries are related to higher-spin fields. 
They are discussed in 
Appendix \ref{sec:HigherSpin}.

%%%%%%%%%%%%%%%%%%%%%%%%%%%%%%%%%%%%%%%%%%%%%%%%%%%%%%%%%%%%%%%%%%%%%%
%%%%%%%%%%%%%%%%%%%%%%%%%%%%%%%%%%%%%%%%%%%%%%%%%%%%%%%%%%%%%%%%%%%%%%

\section{Discussion}
\label{sec:conclusiondiscussion}
\hspace{0.51cm}
In this section, we discuss some issues which 
remain to be clarified. 

\vspace{0.2cm}

We first comment on the massless fields that 
appear in string theory. 
We have considered only the gravity and $U(1)$ gauge field. 
First, 
we have not yet understood how 
anti-symmetric tensor fields arise in this model. 
For the dilaton, one possibility is to introduce it 
as an overall factor:
\begin{equation}
A_{a}= e^{\phi}\left(
ia_{a}^{b}\nabla_{b}+\cdots
\right). 
\end{equation}
Substituting this ansatz into the equation of 
motion (\ref{equationofmotion}), 
we find that only constant solutions solve it. 
As in the case of the equation 
(\ref{equationofmotionRF}), 
we probably need to take account of loop corrections 
in order to obtain non-trivial solutions. 
Second, we assumed that 
there is no torsion 
in subsection \ref{sec:classicalsolutions}. 
Solving the equation of motion (\ref{equationofmotion}) 
with non-zero torsion, 
we find that it propagates as a 
rank-three anti-symmetric tensor field.  
This may provide a clue 
to understanding how anti-symmetric tensor fields appear 
in our model. 
Third, 
although we have considered only the $U(1)$ gauge group,
it does not seem difficult to introduce 
other guage symmetries by extending $G$.

Next, let us discuss what happens if we try to express
the action in terms of the local fields 
that appear in the expansion (\ref{expansionA-x}). 
Because matrices are regarded as differential 
operators defined on a manifold $E_{prin}$,
a naive definition of the trace gives a divergent result.
We need to introduce the reguralizion
in such a way that the cut-off 
corresponds to the matrix size $N$. 
In this case, we can introduce
the heat kernel, using the Laplacian on the principal bundle,
and we could probably use it as the regularization.  
It would be interesting if the action 
can be obtained in a good approximation.

\vspace{0.4cm}

As mentioned in the last paragraph of 
subsection \ref{sec:newmatrixspace}, 
the global supersymmetry is destroyed if we restrict the matrix space. 
There would be two possible ways 
to implement the supersymmetry in our formulation. 
One is to use supergroups as $G$, and the other is 
to use supermanifolds as the base space $M$. 
Here we discuss the latter possibility. 
In this case, the bosonic variable would be expanded as 
\begin{equation}
A_a\sim
ia_a^\mu(x)\nabla_\mu
+\frac{i}{2}\bar{\chi}_a^\alpha(x)\nabla_\alpha, 
\label{expansion_supersymmetric}
\end{equation}
where $\nabla_\mu$ and $\nabla_\alpha$ 
are the covariant derivatives on the superspace. 
If we consider the unitary transformation generated by  
\begin{equation}
\Lambda
=i\bar{\epsilon}^\alpha(x)\nabla_\alpha
+2(\bar{\epsilon}(x)\gamma^\mu\theta)\nabla_\mu, 
\end{equation}
the fields introduced in (\ref{expansion_supersymmetric})
transform according to  
\begin{equation}
\delta a_{a}^{\mu}(x)=2i\bar{\epsilon}(x)
\gamma^{\mu}\chi_{a}(x), \quad 
\delta \chi_{a}^{\alpha}(x)=2\nabla_{a}\epsilon^{\alpha}(x). 
\end{equation}
Thus we can regard $a_a^\mu(x)$ and $\chi_a^\alpha(x)$ 
as the vielbein and the gravitino. 
In this manner, 
it is possible to include the local supersymmetry 
into the unitary symmetry of the matrix model. 
Because the bosonic variable $A_{a}$ 
contains the gravitino, 
there is no reason to consider 
the fermionic variable $\psi_{\alpha}$. 
Therefore there is no guiding principle to determine the 
form of the action. 
If we take 
a Yang-Mills type action, 
it gives a second-order equation for the gravitino, 
although it gives the Einstein equation 
as we have seen in subsection
\ref{sec:classicalsolutions}. 
To pursue this possibility further, 
we should search for a better action.

\vspace{0.4cm}

In Appendix \ref{sec:HigherSpin}, 
we discuss the fact 
that matrix models contain the higher-spin gauge fields 
with corresponding gauge symmetries. 
There remains a problem concerning the fields 
that appear as the coefficients of higher powers of ${\cal O}_{ab}$. 
At first order in ${\cal O}_{ab}$, 
we find that they are higher-spin analogues 
of the spin connection (see appendix \ref{higherspinLL}). 
On the other hand, 
we have not obtained a complete understanding 
for higher orders. 
One of the advantages of our formulation is 
the manifestation of the gauge invariance 
even with the interaction. 
For this reason, it may be interesting to analyze 
higher-spin gauge theories based on 
matrix models. 

In our model, 
higher-spin fields do not have mass terms at the tree level. 
To obtain string theory, 
we should clarify how they acquire masses without explicitly 
breaking higher-spin gauge symmetries. 
If we are able to elucidate 
the origin of the masses and 
resolve the problem of supersymmetry, 
we will get a deeper understanding of
string theory and matrix models.

\section*{Acknowledgements}
The authors would like to thank F. Kubo and K. Murakami 
for helpful discussions. 
The work of M. H and Y. K was supported in part by JSPS Research Fellowships 
for Young Scientists. 
This work was also supported in part by a Grant-in-Aid for
the 21st Century COE ``Center for Diversity and Universality in
Physics''.

\renewcommand{\theequation}{\Alph{section}.\arabic{equation}}
\appendix

\section{Proof of (\ref{irreducibledecompregular})} 
\label{sec:proofofregulardecomposition}
\setcounter{equation}{0} 
\hspace{0.51cm}
In this appendix, we give the proof of (\ref{irreducibledecompregular}). 
We assume the group $G$ to be compact. 
We use 
the orthonormality of the representation matrices expressed as 
\begin{eqnarray}
\frac{1}{Vol(G)}\int dgR_{ij}^{\langle r\rangle}(g)^{\ast}
R_{kl}^{\langle r^{\prime}\rangle}(g)=\frac{1}{d_{r}}
\delta^{\langle r\rangle \langle r^{\prime}\rangle}
\delta_{ik}\delta_{jl}, 
\end{eqnarray}
where $dg$ is the Harr measure and 
$R_{ij}^{\langle r\rangle}(g)$ is the representation matrix 
for the irreducible representation $r$. 
Furthermore, the representation matrices 
$R_{ij}^{\langle r\rangle}(g)$ form a complete set of 
smooth functions from 
$G$ to ${\mathbb C}$. 
Therefore, any smooth function $f(g)$ can be expanded as 
\begin{equation}
f(g)=\sum_{r}c_{ij}^{\langle r\rangle}
\sqrt{d_{r}}R_{ij}^{\langle r\rangle}(g), 
\label{expansionoff(g)}
\end{equation}
where the sum is taken over all irreducible representations. 
We now see how $c_{ij}^{\langle r\rangle}$ 
transform under the action of $G$. 
The right-hand side transforms as 
\begin{eqnarray}
\hat{h}:c_{ij}^{\langle r\rangle}R_{ij}^{\langle r\rangle}(g)
&\rightarrow&
c_{ij}^{\langle r\rangle}R_{ij}^{\langle r\rangle}(h^{-1}g) \cr
&=&
c_{ij}^{\langle r\rangle}R_{ik}^{\langle r\rangle}(h^{-1})
R_{kj}^{\langle r\rangle}(g), 
\end{eqnarray} 
which shows that $c_{ij}^{\langle r\rangle}$ transforms as 
\begin{eqnarray}
c_{ij}^{\langle r\rangle}
\rightarrow
^{T}\!\!R_{ik}^{\langle r\rangle}(h^{-1})c_{kj}^{\langle r\rangle}. 
\end{eqnarray}
The index $i$ of $c_{ij}$ transforms 
as the dual representation 
of $r$, while the index $j$ is invariant. It follows that 
$c_{ij}^{\langle r\rangle}$ represents $d_{r}$ copies of 
the dual representation of $r$. 
Thus we have found that the regular representation is decomposed as 
(\ref{irreducibledecompregular}). 

%%%%%%%%%%%%%%%%%%%%%%%%%%%%%%%%%%%%%%%%%%%%%%%%%%%%%%%%%%%%%%%%%%%%%%%%%%%%%

\section{Speculation on Higher-Spin Gauge Fields}
\label{sec:HigherSpin}
\setcounter{equation}{0} 
\hspace{0.51cm}
In this appendix, we discuss the higher-spin gauge fields 
that appear in the expansion of $A_{a}$ with respect to covariant derivatives. 
\subsection{
Interpretation of $a_{a}^{(\mu_{1}\cdots\mu_{s-1})}$ 
and higher-spin gauge symmetries}
\hspace{0.51cm}
As we have seen from the transformation law under the differmorphism, 
the fields $a_{a}^{\mu_{1}\cdots\mu_{s-1}}$ transform as 
rank-$(s-1)$ totally symmetric ternsor fields 
(see (\ref{diffforspinthree})). 
The field $a_{a}^{\mu}(x)$ is identified with the vielbein, 
which can be regarded as a gauge field 
associated with the symmetry generated by
$\Lambda\sim i\lambda^{\mu}(x)\partial_{\mu}$. 
Therefore, it is natural to regard
$a_{a}^{\mu_{1}\cdots\mu_{s-1}}(x)$ 
as a gauge field associated with
\begin{equation}
\Lambda=\frac{i^{s-1}}{2}\{\lambda^{\mu_{1}\cdots \mu_{s-1}}(x),
\partial_{\mu_{1}}\cdots \partial_{\mu_{s-1}} \}. 
\label{generatorofhigherspin}
\end{equation}
We first examine the inhomogeneous term 
in such a transformation. It comes from the commutator 
between this $\Lambda$ and the term
$\frac{i}{2}\{a_{a}^{\mu}(x),\partial_{\mu}\}$ in $A_{a}$. 
Indeed, we can show that 
the field $a_{a}^{\mu_{1}\cdots\mu_{s-1}}(x)$ transforms 
according to 
\begin{eqnarray}
\delta a_{a}^{\mu_{1}\cdots\mu_{s-1}}(x)=a_{a}^{\nu}
\partial_{\nu}\lambda^{\mu_{1}\cdots\mu_{s-1}}(x)+\cdots,
\end{eqnarray}
which is a natural generalization of 
(\ref{U(1)gaugesymm}) and (\ref{diffvielbein}). 

We next discuss the homogeneous terms. 
The higher order derivatives in 
$\Lambda$ given in (\ref{generatorofhigherspin})
cause notrivial complexity compared 
to the case of the lower-spin gauge symmetries. 
Fields with different spins mix 
in a complicated manner, 
and also higher-order derivative terms appear. 
The latter 
implies the existence of higher-order derivative interactions.

\subsection{Interpretation of $a_{(1)}{}_{a}^{b\cdots,cd}$ 
and higher-spin analogue of local Lorentz symmetry}
\label{higherspinLL}
\hspace{0.51cm}
It is natural to generalize the relation between 
the vielbein and the metric, 
\begin{equation}
g_{\mu\nu}(x)=a_{\mu}^{a}(x)a_{\nu}^{b}(x)\delta_{ab}, 
\label{metricandvielbein}
\end{equation}
to higher-spin fields as
\begin{equation}
\phi^{\mu_{1}\cdots\mu_{s}}(x)=
a_{a}^{(\mu_{1}}(x)a_{b}^{\mu_{2}\cdots\mu_{s})}(x)\delta^{ab}. 
\label{HSmetriclikefieldfromvielbeinlikefield}
\end{equation}
Here $\phi^{\mu_{1}\cdots\mu_{s}}(x)$ 
is a rank-$s$ totally symmetric tensor field \cite{Fronsdal,deWitFreedman} 
which transforms according to 
\begin{eqnarray}
\delta \phi^{\mu_{1}\cdots\mu_{s}}(x)=
\partial^{(\mu_{1}}\lambda^{\mu_{2}\cdots\mu_{s})}(x)+\cdots
\label{higherspingaugetransformation}
\end{eqnarray}
under the transformation generated by (\ref{generatorofhigherspin}). 
There are some studies of 
higher-spin gauge theories based on the relation 
(\ref{HSmetriclikefieldfromvielbeinlikefield}). 
See \cite{hep-th/0503128} for reviews. 

We now recall the role of the local Lorentz symmetry. 
It removes 
the anti-symmetric part of the vielbein $a_{\mu}^{a}(x)$, and 
the remaining symmetric part corresponds to 
the metric ternsor. 
We find that a mechanism similar to the
higher-spin gauge fields $a_{a}^{(\mu_{1}\cdots\mu_{s-1})}$
exists in our model. 

We consider the case of spin three ($s=3$) as an example. 
Because the ordinary local Lorentz symmetry is generated by 
$\Lambda\sim i\lambda_{(1)}^{ab}{\cal O}_{ab}$, 
it is natural to consider 
\begin{equation}
\Lambda=\frac{1}{4}
i\{i
\{\lambda_{(1)}{}^{a, bc},{\cal O}_{bc}\},\partial_{a}\}. 
\label{generatorofhsLL}
\end{equation}
Then 
$a_{a}^{\mu\nu}(x)$ transforms as
\begin{eqnarray}
\delta a_{a}^{\mu\nu}(x) \sim 
-\lambda_{(1)}{}^{(\mu,}{}_{ab} (x)a_{b}^{\nu)}(x), 
\end{eqnarray}
where $\lambda_{(1)}^{\mu,ab}(x)=\lambda_{(1)}^{c,ab}(x)e_{c}^{\mu}(x)$. 
This is a generalization of (\ref{LLforbackgrondvielbein}). 
In this case, however, we have a mixing of fields with different spins 
($a_{a}^{\mu\nu}(x)$ and $a_{b}^{\nu}(x)$) 
due to the existence of the derivative in (\ref{generatorofhsLL}). 
Under this transformation, 
the spin-three field $\phi^{\mu\nu\rho}(x)$ is invariant: 
\begin{eqnarray}
\delta \phi^{\mu\nu\rho}(x)\rightarrow 
a_{a}^{(\mu}(x)\delta a_{b}^{\nu\rho)}(x)\delta^{ab}
\sim -\lambda_{(1)}{}^{(\nu,}{}_{ab}(x)a_{a}^{\mu}(x)a_{b}^{\rho)}(x)
=0, 
\end{eqnarray}
where we have used 
$\lambda_{(1)}{}^{\mu, ab}(x)=-\lambda_{(1)}{}^{\mu, ba}(x)$. 
This symmetry
removes unnecessary components from the field 
$a_{a}^{\mu\nu}(x)$ 
and leaves the field $\phi^{\mu\nu\rho}(x)$ as
the dynamical degrees of freedom. 

Because the spin connection appears in $A_{a}$ as the term 
$i\{a_{(1)}{}_{a}^{bc},{\cal O}_{bc}\}$, 
we expect that
a term like
$\{\{a_{(1)}{}_{a}^{b,cd},{\cal O}_{cd}\},\partial_{b}\}$
plays an analogous role as the spin connection. 
Actually, we can check that $a_{(1)}{}_{a}^{b,cd}$ 
transforms inhomogeneously under the transformation 
(\ref{generatorofhsLL}) as 
\begin{equation}
\delta a_{(1)}{}_{a}^{b,cd}=\partial_{a}\lambda_{(1)}{}^{b,cd}+\cdots . 
\end{equation}
A similar mechanism works also for $s\ge 4$.

In this manner, higher-spin sectors are included in our 
formulation. 
However, the positivity of the theory is still unclear. 
We have not yet discussed fields appearing 
with the second order in ${\cal O}_{bc}$, such as 
$a_{(2)}{}_{a}^{bc,de}(x)$. 
In such cases, 
analyses become difficult, 
because the higher-spin gauge symmetries mix fields 
with different spins in a complicated manner.

\end{document}